\documentclass[sigconf, authorversion, screen]{acmart}
\usepackage{threeparttable}
\usepackage{multirow}
\usepackage{enumitem}
\usepackage{subcaption}
\usepackage{mathtools}
\usepackage{bbm}
\usepackage{makecell}
\usepackage{subcaption}
\usepackage{amsmath}
\usepackage{diagbox}
\usepackage{hyperref}
\DeclareMathOperator*{\argmin}{argmin} 
\usepackage[linesnumbered,ruled]{algorithm2e}
\SetAlgorithmName{Procedure}{Procedure}{List of Procedures}
\AtBeginDocument{%
  \providecommand\BibTeX{{%
    \normalfont B\kern-0.5em{\scshape i\kern-0.25em b}\kern-0.8em\TeX}}}

\copyrightyear{2022}
\acmYear{2022}
\setcopyright{acmcopyright}\acmConference[PROMISE '22]{Proceedings of the 18th
International Conference on Predictive Models and Data Analytics in Software
Engineering}{November 17, 2022}{Singapore, Singapore}
\acmBooktitle{Proceedings of the 18th International Conference on Predictive Models
and Data Analytics in Software Engineering (PROMISE '22), November 17, 2022,
Singapore, Singapore}
\acmPrice{15.00}
\acmDOI{10.1145/3558489.3559067}
\acmISBN{978-1-4503-9860-2/22/11}

\begin{document}

\title[Measuring design compliance using neural language models] {Measuring design compliance using neural language models -- an automotive case study}


\author{Dhasarathy Parthasarathy, ~Cecilia Ekelin}
\affiliation{%
  \institution{Volvo Group, Sweden}
}

\author{Anjali Karri, ~Jiapeng Sun, ~Panagiotis Moraitis}
\affiliation{%
  \institution{Chalmers University of Technology, Sweden}
}








\renewcommand{\shortauthors}{Parthasarathy, et al.}

\begin{abstract}
  As the modern vehicle becomes more software-defined, it is beginning to take significant effort to avoid serious regression in software design. This is because automotive software architects rely largely upon manual review of code to spot deviations from specified design principles. Such an approach is both inefficient and prone to error. In recent days, neural language models pre-trained on source code are beginning to be used for automating a variety of programming tasks. In this work, we extend the application of such a Programming Language Model (PLM) to automate the assessment of design compliance. Using a PLM, we construct a system that assesses whether a set of query programs comply with \emph{Controller-Handler}, a design pattern specified to ensure hardware abstraction in automotive control software. The assessment is based upon measuring whether the geometrical arrangement of query program embeddings, extracted from the PLM, aligns with that of a set of known implementations of the pattern. The level of alignment is then transformed into an interpretable measure of compliance. Using a controlled experiment, we demonstrate that our technique determines compliance with a precision of 92\%. Also, using expert review to calibrate the automated assessment, we introduce a protocol to determine the nature of the violation, helping eventual refactoring. Results from this work indicate that neural language models can provide valuable assistance to human architects in assessing and fixing violations in automotive software design.\looseness=-1
\end{abstract}



\begin{teaserfigure}
    \includegraphics[width=1.0\linewidth]{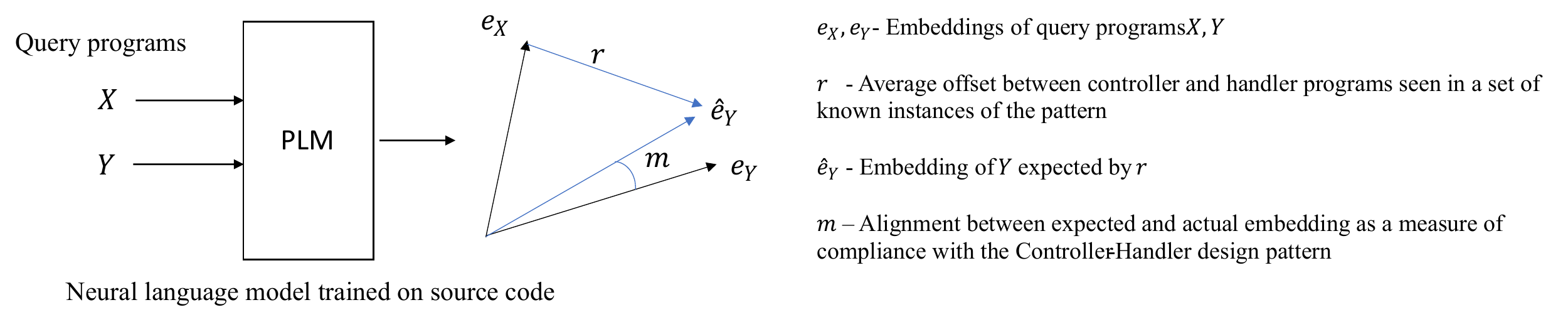}
    \caption{Overview: We demonstrate a system that automatically assesses whether query programs $(X, Y)$ complies with the automotive \emph{Controller-Handler} design pattern. The heart of the system is a neural language model pre-trained on source code. The assessment process compares geometrical properties of the embeddings of query programs with that of a set of known instances of the pattern. The comparison is then converted into a score that allows architects to interpret the level of compliance.\looseness=-1}
    \label{fig:overview}
  \end{teaserfigure}
\maketitle

\section{Introduction}
'If you think good design is expensive, try bad design', goes the aphorism. While this observation can headline any design effort, it is certainly a prime motivator in the design of software. Very generally, the process of software design attempts to envision a software solution that meets a given set of requirements \cite{ISO19759}. The classic design process achieves this using a combination of tools including design principles, architecture models, interface specifications, etc., that parallelly guide, if not instruct top-down, the implementation of the solution. Meeting core business requirements may be its primary objective, but software design often aspires further to address several non-functional concerns to increase the likelihood that the solution operates and evolves sustainably \cite{DBLP:journals/software/ChenBN13}. The expanded set of concerns inevitably complicates the design process, which now becomes an act of trading-off concerns in multiple dimensions, under the shadow of constant uncertainty. In recent years, neural language models pre-trained on large source code corpora have started becoming building blocks for automating a variety of complex programming tasks like code completion and program repair (\cite{DBLP:conf/icse/HindleBSGD12} and \cite{DBLP:conf/oopsla/PuNSB16}, for example). If such programming tasks, which often require nuanced judgment, can be automated, can a similar approach be applied to automate design tasks? In this work, we take initial steps towards answering this question by investigating a use case in automotive software design. \looseness=-1

\vspace{1mm}
\noindent \textbf{Need for assistance in automotive software design} -- The modern vehicle is increasingly software driven. Software plays a central role in realizing a variety of in-vehicle functions like preventive safety, driver assistance, energy management, etc. Not only is increasing the footprint of software essential to meet the growing demand for functionality, vehicle manufacturers are increasingly realizing that well-designed software is a key requirement to meet this demand sustainably \cite{DBLP:conf/icse/MagnussonLL18}. However, there are several factors that complicate the design and evolution of automotive software. A strict regulatory environment, decades of legacy, complex integration chains, the strong influence of non-functional concerns like safety and security, and strong hardware coupling are prominent among them. Moreover, since the automotive industry has its roots in traditional disciplines like mechanical and electrical engineering, knowledge of principles and practices of software engineering is less widespread. Therefore, delivering software at high cadence, while minimizing design compromises and preventing major threats to sustainable evolution of the code base, remains a formidable challenge. Currently, the industry relies upon experienced software architects to manually assess the code for design violations and intervene with changes when necessary. With violations from specified design being inevitable in practice, one advantage of manual review is that the expert is able to exercise nuance and judgment on whether a given violation is acceptable. The disadvantage, of course, is that manually assessing thousands of lines of code is effort-intensive. The intensity of effort alone increases the likelihood that major violations are left undiscovered and the overall design regresses, with harmful consequences for system evolution. Automatic assessment of design, with levels of nuance comparable to a human expert, would therefore provide vital assistance in increasing the speed and effectiveness of design intervention. \looseness=-1

\vspace{1mm}
\noindent \textbf{Neural language models for software design} --
The application of neural language models for automating programming tasks is fundamentally based upon the naturalness hypothesis \cite{DBLP:journals/csur/AllamanisBDS18}, which recognizes that software is a form of communication. Neural Programming Language Models (PLMs), pre-trained on code corpora, exploit such infused elements of human communication to learn a statistical model of programming. Such knowledge lies at the foundation of its ability to automate complex programming tasks. In our attempt to apply PLMs for automating design-related tasks, we start by considering whether design information is also naturally communicated in code. Generally, programmers choose to augment self-explanation in code so that fellow-programmers find it easy to extend. A basic explanatory technique like using well-worded program statements, in a clearly evident sequence, accompanied by lucid natural language comments clearly helps code extension in relatively local scopes. In parallel, carefully wording and characterizing entities like methods, modules, or classes, and the ways in which they relate, interact, and are packaged, promote more global extension. Infusing such explanation, which is largely complementary to program logic, clearly achieve many of the same objectives of a top-down design exercise. In fact, the co-evolution of design and solution -- the 'code as design' approach -- is itself a natural byproduct of using high-level PLs \cite{reeves1992software}. Put simply, with software naturally containing algorithmic and explanatory channels, the latter is likely to include information about design. Thus, irrespective of whether it emerges bottom-up as a result of programming or top-down as a result of a design process, \emph{elements of software design occur naturally in source code}. Given that (1) PLMs successfully understand statistical properties of natural programming, and (2) elements of design occur naturally in source code, we reason that \emph{PLMs pre-trained on large code corpora are likely to understand elements of design}. The purpose of this study is to both verify this reasoning and exploit its potential.\looseness=-1

\vspace{1mm}
\noindent \textbf{Problem statement} -- We envision a system $\mathcal{S}$ that uses a PLM $\mathcal{F}$ to assess whether a set of query programs/files $Q$, drawn from a corpus $\mathcal{Q}$, complies with a design pattern $\mathcal{D}$ specified for the corpus. A score $m$ calculated by the system provides a measure of compliance. \looseness=-1

\begin{equation}
    \label{eq:dp_assessment}
    \begin{aligned}
        m &= \mathcal{S} (Q, ~\mathcal{D}; ~\mathcal{F}), ~ Q \subseteq \mathcal{Q}        
    \end{aligned}    
\end{equation}

To construct and evaluate such a system, we pose the following research questions.

\begin{itemize}[wide, label=, labelindent=-5pt]
    \item \textbf{RQ1} -- Can the system $\mathcal{S}$ for assessing design compliance be constructed using a neural language model trained on code?
    \item \textbf{RQ2} -- Does the assessment improve when the PLM is explicitly provided with information relevant to design pattern $\mathcal{D}$?
    \item \textbf{RQ3} -- Can the measure $m$ be communicated in a way that makes it easy for an architect to understand the compliance of $Q$ with $\mathcal{D}$? \looseness=-1        
\end{itemize}

Results from our study\footnote{We release the implementation of our compliance assessment system \href{https://github.com/dhas/measuring-design-compliance.git}{here}. Since the test corpus $\mathcal{Q}$ is proprietary, we include examples that illustrate its content.} show that it is indeed possible to construct such a system for measuring design compliance. Such a neural language modeling approach to automatically assess design compliance has the potential to improve the chances of quickly identifying (and subsequently correcting) design violations, thus promoting sustainable evolution of the code base with increased cadence.
\section{Choosing a corpus and design pattern for study}
\label{sec:dp}
We now describe (1) the corpus $\mathcal{Q}$, from which query programs are drawn, and (2) the pattern $\mathcal{D}$ against which their design is assessed. \looseness=-1

\begin{figure}[h]
    \centering
    \includegraphics[width=0.9\linewidth]{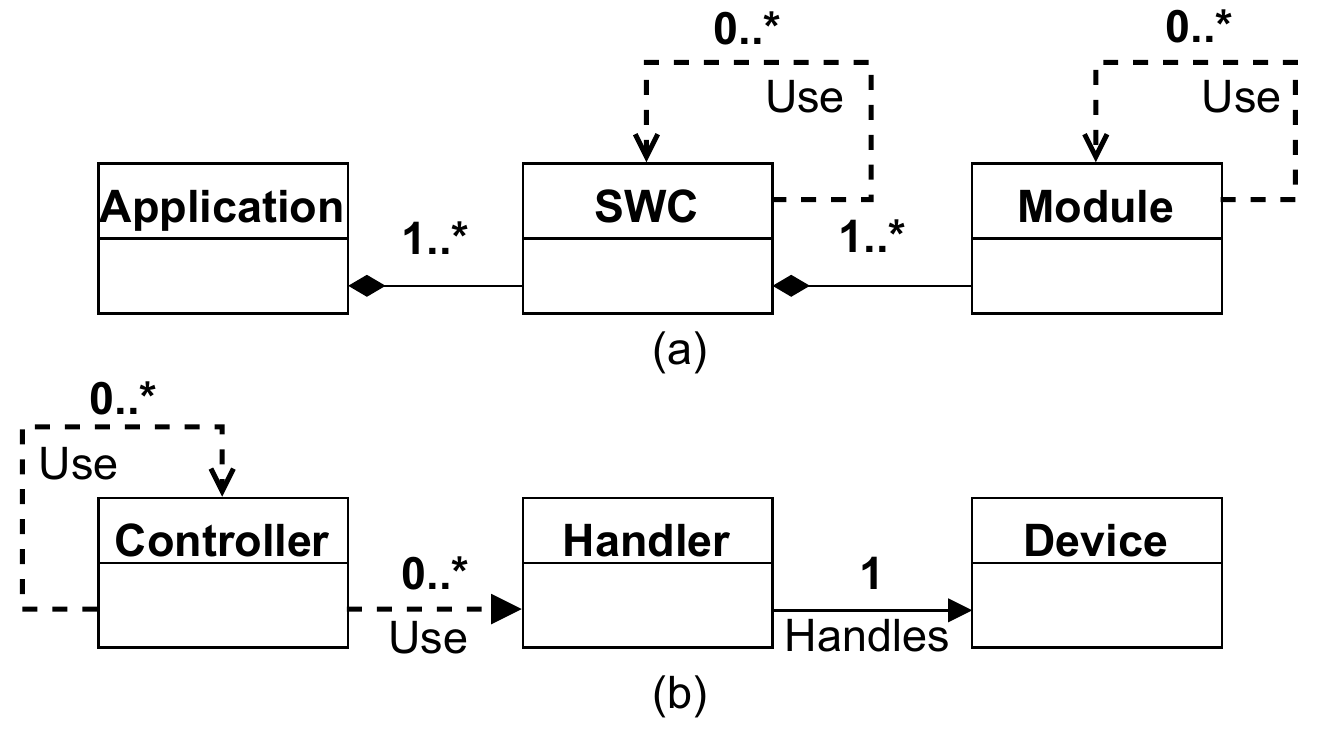}
    \caption{(a) SWCs collaborating to implement a vehicle function, (b) Controller-Handler software design pattern for automotive control systems}
    \label{fig:swc-app}
\end{figure}

\vspace{1mm}
\noindent \textbf{Truck Application Software corpus} -- In a modern vehicle, the overall system which governs the automatic control of in-vehicle functions is generally referred to as the Electrical/Electronic (E/E) system (\cite{2019-01-0862}). In this system, the basic unit of (electronic) hardware is the Electronic Control Unit (ECU). It is typically a microcontroller-based platform that brings together the necessary elements of automatic control -- the control logic, sensors, actuators, and related I/O. Software -- deployed on ECUs to realize the control logic -- is our focus here. For this study, we use Truck Application Software (TAS), a corpus of $\sim$5k files of C-language code, that implements in-vehicle functionality for the Volvo Group's truck platforms. Principles of software design adopted in TAS stem mainly from the Automotive Software Architecture (AUTOSAR) industry standard \cite{DBLP:journals/insk/Bunzel11}. The basic unit of software defined by AUTOSAR is the Software Component (SWC), which is an independently deployable unit of functionality. At the core of each SWC is a set of C files, called software modules, which collectively realize the functionality of the SWC. Upon deployment, SWCs interact with each other to collectively realize control applications (Figure \ref{fig:swc-app}a). The decomposition of control logic into a set of SWCs is fundamental for achieving several objectives of automotive software design including the management of complexity, separation of concerns, and the promotion of reuse. Therefore, the notion of a SWC is, de facto, also the basic tool for software design in TAS. Any design pattern of increased sophistication adopted in this corpus, builds upon the idea SWCs. One such design pattern, which we focus upon in this study, is described below. \looseness=-1

\vspace{1mm}
\noindent \textbf{The Controller-Handler design pattern} -- Let us now consider an example application\footnote{Refer to roofHatch in the released code for an illustration} in TAS -- roof hatch control -- and its design. Trucks are sometimes equipped with a hatch on the roof, which the driver can control to adjust the flow of air and the amount of ambient light. The hatch is equipped with motors that effect this control. One design principle, used in TAS, to implement such a function is the separation of the core logic for hatch adjustment, the \emph{Controller}, from the logic that handles the motors, the \emph{Handler}. The main reason for this separation is that the roof-hatch motors are physically wired to specific hardware pins in a specific ECU. This means that the handler needs to be deployed on this particular ECU and use the designated pins to control the motor. In contrast, the application logic in the controller is not bound to a specific ECU, which allows more freedom in deciding where it can be deployed. Since automotive ECUs (traditionally) are quite resource constrained, this \emph{Controller-Handler} (CH) pattern offers a way to efficiently utilize the available resources. Moreover, such hardware abstraction is essential to make cost-effective product offerings. A truck typically needs to support a high level of product configuration to be able to fit a variety of transport operations and market segments. A roof hatch control application that properly implements CH can help offer trucks with different variants of roof-hatch motors, each tuned to meet a certain customer demand. Having separated the application logic, the controller can be reused over all these variants. Due to its prevalence in TAS we focus on the CH design pattern in this study. Formally (see Figure \ref{fig:swc-app}b), the CH pattern advocates the implementation of an in-vehicle control application using a set of SWCs $P = \{C, H_1, H_2, ... H_N\}$. Here, the \emph{Controller} component $C$, implements the core control logic, while \emph{Handler} components $H_i, i=1...N$ implement hardware-specific logic. In practice, since the handler components are usually independent of each other, the CH design pattern can be defined as applying to each pair $P = (C, H_i)$ of controller and handler SWCs used to realize the overall application. Apart from roof hatch control, applications in TAS that adopt this design pattern include washer and wiper control, exterior lights control and accelerator pedal control. \looseness=-1

While the reasoning behind the CH pattern is intuitive, compliance is not always achievable in practice. For instance, the responsibility split between the controller and handler must be at the right level to, among other things, avoid duplication of logic across handler variants. Roof hatch control includes special logic to ensure electrical safety during actuation. Placing all of this safety logic in the handler (and not just motor-specific parts) leads to duplication. If there happens to be a violation in some handler variant which implements more safety logic than necessary, spotting this is not easy for an architect who was not intimately involved in its design. On the other hand, there may be sufficient clues in tokens used in the compromised handler source code that a neural PLM may find anomalous.\looseness=-1

\section{Constructing a system for assessing design compliance}
\label{sec:compliance}
Having fixed the query corpus $\mathcal{Q}$ as TAS and the design pattern $\mathcal{D}$ as Controller-Handler for the study, we restate our objective. We aim to construct a system $\mathcal{S}$ that assesses whether an ordered pair $Q = (X, Y), ~ X, Y \in \mathcal{Q}$ of SWCs comply with the Controller-Handler design pattern. Though either of the SWCs in this pair can be realized using multiple software modules (or programs), at this point it is simpler to consider the case where each SWC is realized as one program. We relax this condition at a later point. The following sections describe the process of constructing the compliance assessment system $\mathcal{S}$ that we envision in Eq.\ref{eq:dp_assessment}. \looseness=-1

\vspace{1mm}
\noindent \textbf{Pre-training a PLM} -- In this work, we consider a program $X = (t_1, t_2, ..., t_N)$ to be a source code file containing a sequence of tokens. We then define a PLM to be a language representation model of the form $\mathcal{F}(X_M) \rightarrow X$ pre-trained as a masked language model, first introduced in BERT \cite{DBLP:conf/naacl/DevlinCLT19}. The core task in pre-training such a model is Masked Reconstruction (MR), shown in Eq.\ref{eq:obj-mr}\footnote{In practice, a differentiable cross-entropy loss is used}. In this task, the PLM is provided a masked program $X_M$, which produced by replacing a fraction of tokens in $X$ with a mask token $\mathbf{t}$. The model is then tasked to recover tokens in masked positions, as a result of which it learns contextual meanings of programs.
\begin{equation}
    \label{eq:obj-mr}
    \begin{aligned}
    MR(X; \mathcal{F}) &= ~\mathcal{F}(X_M)[j] == X[j], \\
        j &= \{i:t_i=\mathbf{t}, t_i \in X_M\}   
    \end{aligned}    
\end{equation}
Since our aim is to assess design compliance in TAS which is a C-language corpus, we pre-train a monolingual PLM on C code. As pre-training corpus $\mathcal{P}$, we use $\sim$75M files of C code derived from the GitHub public dataset\footnote{https://console.cloud.google.com/marketplace/details/github/github-repos}. The model is then pre-trained by minimizing the objective shown in Eq. \ref{eq:obj-mlm}. Prior to being fed into a PLM, each program is tokenized and split further into a sequence of subwords using the Byte Pair Encoding (BPE) \cite{DBLP:conf/acl/SennrichHB16a} mechanism.
\begin{equation}
    \label{eq:obj-mlm}
    \begin{aligned}
    \mathcal{F} &:= \argmin_{\mathcal{F}} ~~~\mathbb{E}_{X \in \mathcal{P}} ~~~MR(X; ~\mathcal{F})
    \end{aligned}    
\end{equation}
The procedure described in Eq.\ref{eq:dp_assessment} assesses whether a set of programs $Q$ complies with a design pattern $\mathcal{D}$. Practically, however, feeding an entire program into the PLM is an issue because C programs tend to be long. The average length of a C program is $\sim$5k subwords in the GitHub corpus and $\sim$7k subwords in the TAS corpus. The Transformer architecture \cite{DBLP:conf/nips/VaswaniSPUJGKP17}, which is the mainstay of several previously reported foundational PLMs like CodeBERT \cite{DBLP:conf/emnlp/FengGTDFGS0LJZ20}, is typically configured to handle input sequences of length 512-1024. This is because the vanilla self-attention mechanism is of quadratic compute and memory complexity, which makes it impractical for longer input sequences. To be able to assess long programs, we therefore base the PLM $\mathcal{F}$ on the more efficient Reformer \cite{DBLP:conf/iclr/KitaevKL20} architecture. Combining  locality-sensitive-hashing and reversible residual layers, the Reformer handles long sequences much more efficiently. By configuring the input sequence length to 8192, we are able to feed around 80\% of programs in the TAS corpus into the Reformer-based $\mathcal{F}$ intact with manageable memory and computational complexity. Longer programs are truncated to this length. The Reformer encoder $\mathcal{F}$ of $\sim$180M parameters with 6 self-attention layers (each with 8 heads) was trained from scratch on 16 Nvidia Tesla V100 GPUs until the MR accuracy on a validation set of 5k files reached 95.12\%. \looseness=-1

\vspace{1mm}
\noindent \textbf{Assessing compliance by manual review} -- Recalling the CH design pattern described in Section \ref{sec:dp}, let us now consider how a human architect would assess whether a query pair $(X, Y)$ of programs complies with this pattern. The architect would normally do this by reviewing the code (or the `naturalness') of the programs and assess whether $X$ and $Y$ respectively embody core principles of a controller and its associated handler. It is, however, important to note that the CH pattern defines expectations jointly on the pair and not on individual programs. Therefore, a starting point for the architect would be to juxtapose related parts from the pair (sometimes mentally) and then conduct the assessment. We find it useful to refer to such a juxtaposition as $XY$ -- the `jointness' of the two programs. It is on this -- at times abstract -- representation XY that the architect assesses whether principles of the CH pattern are complied with. In the example of roof hatch control, signs of compliance include that the interface for the handler is a pure abstraction of the hardware interface for the hatch motors. That is, the handler does not contain extra logic e.g. to protect the motors from over usage. That kind of logic would be part of the controller. Similarly, the controller should make use of the handler interface only to interact with the motors. Signs of deviation from the pattern are the opposite of what has been described. That is, the handler contains too much control logic or the controller interacts directly with the motors. Not only is manually assessing the jointness $XY$ for signs of deviation difficult, there are several factors that complicate the process further. First, any instance of the CH pattern is certain to contain code that falls outside the purview of the pattern itself. Hence, an architect will have to identify and assess tenets of the pattern in a diluted context. Second, as a relatively loose pattern, it can be realized in several styles. An architect would therefore need to judge whether a given style of implementation is legitimate. Third, it is practically difficult to construct an ideal realization against which the query programs can be assessed. Usually, the architect relies on a subjective mental model of the pattern, which is not only difficult to explicitly state, but also affects the objectivity of the assessment. Addressing these concerns requires nuanced judgment, which is precisely what a human expert applies. In using a PLM as an alternative to a human expert, we now describe how we address some of these concerns. \looseness=-1

\vspace{1mm}
\noindent \textbf{Assessing compliance using program embeddings} -- The main tool we use for PLM-based compliance assessment is the \emph{program embedding} $e_X$, which is a vector representation of the program $X$ that reflects its semantic properties. As shown in \cite{DBLP:conf/naacl/PetersNIGCLZ18}, there are different ways to extract embeddings from contextual language models, each capturing different aspects of information. After some trial and error, we empirically decide to use the normalized output of the final (6th) layer of $\mathcal{F}$ as shown below. \looseness=-1

\begin{equation}
    \label{eq:embedding}
    \begin{aligned}
        e_X &= \frac{\mathcal{F}_6 (X)} {||\mathcal{F}_6 (X)||}
    \end{aligned}    
\end{equation}

The PLM $\mathcal{F}$ is pre-trained on the masked reconstruction task on millions of program examples. It is therefore reasonable to expect that the embedding $e_X$ is a fairly robust representation of the program $X$ and is insensitive to minor semantic variations. Thus, the process of assessing whether $(X, Y)$ complies with the CH pattern is done, not in the code space, but in a vector space using embeddings $(e_X, e_Y)$. While this pair of embeddings sufficiently represent the programs individually, an additional representation is needed to address the joint perspective $XY$. One simple model to capture the jointness of a pair of programs would be the offset between their embeddings. \looseness=-1
\begin{equation}
    \label{eq:offset}
    \begin{aligned}
        r_{XY} &= e_Y - e_X
    \end{aligned}    
\end{equation}

Should there exist a benchmark vector $r$ that captures the required level of jointness as prescribed by the CH design pattern, then the assessment of design compliance reduces to checking the alignment between $r_{XY}$ and $r$ in the embedding space. Put otherwise, if $r$ serves as an effective offset vector between the embeddings of the pair of programs $(X, Y)$, i.e., if Eq.\ref{eq:ideal_offset} is satisfied, then this pair comes close to realizing the principles specified by the CH pattern. \looseness=-1
\begin{equation}
    \label{eq:ideal_offset}
    \begin{aligned}
        \hat{e}_Y := e_X + r \approx e_Y
    \end{aligned}    
\end{equation}

As noted earlier among concerns in manual assessment, there is no easy way to construct an ideal realization of the CH pattern in the code space. This means that access to its embedding equivalent $r$ is equally difficult. As a practical alternative, we assess compliance with the \emph{average} realization of the CH pattern, extracted from a set of known instances. That is, given a set $V = \{(C, H)\}_{i=1}^N$ from the TAS corpus that are known to implement the CH pattern, we define a benchmark of average jointness (Eq. \ref{eq:avg_offset}), that averages offset vectors from pairs in $V$. If this benchmark serves as an effective offset for query programs $(X, Y)$, satisfying Eq. \ref{eq:ideal_offset}, then this pair comes close to realizing the average implementation of the CH pattern seen in $|V|$ known instances. Apart from being an intuitive and practical benchmark, by pooling common traits from known instances, $r$ provides a stronger signature for the CH pattern compared to individual instances, where signatures of the pattern are likely to be diluted. \looseness=-1

\begin{equation}
    \label{eq:avg_offset}
    \begin{aligned}        
        r := \frac{1}{|V|} \sum_{(C, H) \in V} e_{H} - e_{C}
    \end{aligned}    
\end{equation}

\begin{figure}[h]
    \centering
    \includegraphics[width=0.5\linewidth]{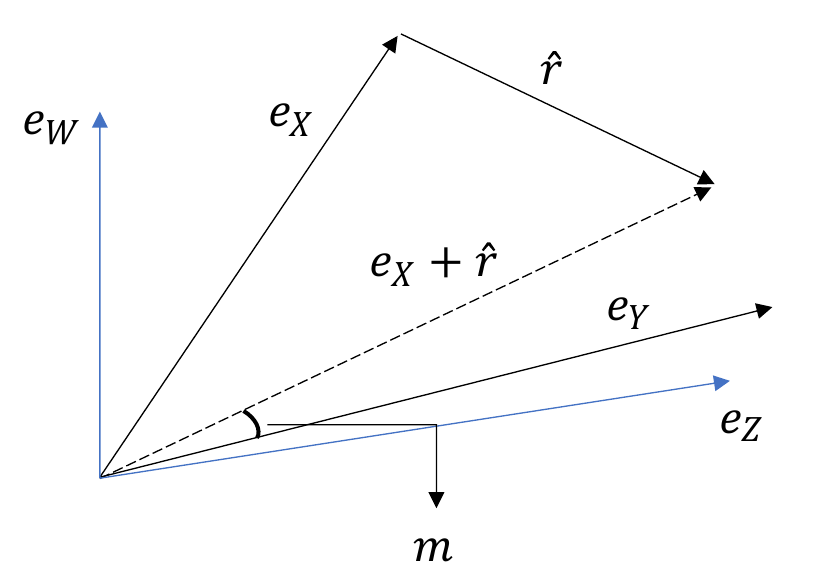}
    \caption{The alignment between the actual handler embedding $e_Y$ and the predicted one $e_X + r$ reflects compliance. Vectors $e_W, e_Z$ illustrate embeddings of programs $W, Z \in \mathcal{Q}$} \looseness=-1
    \label{fig:embedding-rhat}
\end{figure}

As shown using an example in Figure \ref{fig:embedding-rhat}, serving as an offset vector from $e_X$, if $r$ is able to predict a handler embedding $\hat{e}_Y$ that is reasonably close to its actual counterpart $e_Y$, programs $(X, Y)$ are likely to comply with the CH pattern. Such closeness between $\hat{e}_Y$ and $e_Y$ is easily measurable using the cosine similarity between these two vectors. With this method, the assessment system for the CH design pattern $\mathcal{D}$, originally envisioned as Eq.\ref{eq:dp_assessment}, can be re-written as follows. \looseness=-1

\begin{equation}
    \label{eq:avg_assessment}
    \begin{aligned}
        m = \mathcal{S}((X, Y), \mathcal{D}; ~\mathcal{F}, V) = \frac{e_{Y}\cdot (e_{X} + r)} {||e_{Y}||_2 ~~||e_{X} + r||_2}
    \end{aligned}    
\end{equation}

Using cosine similarity as the metric measure -- standard practice for comparing language model embeddings -- results in $-1 \leq m \leq 1$. Then, $m \approx 1$ means that the predicted handler embedding $\hat{e}_Y$ closely aligns with that of the actual handler $e_Y$, indicating compliance. Thus, as a way to assess compliance with the CH design pattern, we substitute a complex code review process with a vastly simpler comparison of embeddings extracted from a neural language model. \looseness=-1

\vspace{1mm}
\noindent \textbf{Easing interpretation of compliance} -- With cosine similarity, while it is clear that $m = 1$ and $m=-1$ indicate perfect compliance and non-compliance respectively, such perfect scores are rare. Scores in between, which are most likely in practice, are difficult to interpret. In order to provide intuitive human-readable assessment, we convert similarity $m$ into a \emph{rank} $k$. The discrete rank $k$ means that the predicted handler embedding is the $k^{th}$ most similar to that of the actual handler, when compared to the embeddings of all other programs in the TAS corpus. The best indicator of compliance is a rank of $k=1$ when, among all programs in the TAS corpus $\mathcal{Q}$ (excluding the controller $X$) there is no better handler than $Y$ for the controller $X$, as assessed by the benchmark $r$. Conversely, a rank of $|\mathcal{Q}|-1$ means that the predicted embedding is least similar and any other program in the TAS corpus is a better handler than $Y$. This is the worst indicator of compliance. While the rank may be a more interpretable measure, its value is now dependent upon the spread of embeddings around $e_Y$. In the example shown in Figure \ref{fig:embedding-rhat}, even if the prediction is reasonably good, it is of rank $k = 2$, since there is another program $Z \in \mathcal{Q}$, whose embedding is closer to that of the actual handler program $Y$. If there is considerable clustering in the close neighborhood of $e_Y$, then even a good prediction is unlikely to result in a rank close to $1$. We therefore use a simple rule of thumb, where if the predicted embedding lies within 10\% of embeddings most similar to $e_Y$, we define the assessment $l = $ \texttt{True} that the query $(X, Y)$ complies with the CH pattern. If the predicted embedding lies among those of 90\% of the least similar programs, we label the pair as non-compliant. The discrete rank $k$, in addition to a true/false binary assessment of compliance $l$, eases human comprehension of our PLM-based process of assessing design compliance. The complete process of compliance assessment is described in Procedure \ref{alg:compliance}. \looseness=-1

\begin{algorithm}
    \caption{Compliance assessment system $\mathcal{S}$}
    \label{alg:compliance}    
    \SetKwInOut{Parameters}{Parameters}
    \SetKwProg{Fn}{Function}{:}{}
    \SetNoFillComment
    \begin{small}
        \Parameters{Test input $(X, Y)$, PLM $\mathcal{F}$, TAS corpus $\mathcal{Q}$, known instances of the CH pattern $V$}
        \Fn{$M(e_A, ~~e_B)$}{
            $m = \frac{e_A . e_B}{||e_A||_2 ~~ ||e_B||_2}$ \\
            return $m$
        }
        \Fn{$\mathcal{S}(X, Y; \mathcal{F}, V)$}{
            \tcc{Note: $e_X = \mathcal{F}_6(X) / ||\mathcal{F}_6(X)||$}
            $r = \frac{1}{|V|} \sum_{(C, H) \in V} e_{H} - e_{C}$ \\
            $c = \lbrack M(e_Z, ~e_{X} + r) : ~Z \in ~\mathcal{Q} ~\backslash ~\{X\} \rbrack$ \\            
            $k =$ indexof (sort$(c), ~M(e_{Y}, ~e_{X} + r)$) \tcp*[f]{rank}\\             
            $l = k \leq 0.1*|\mathcal{Q}|$  \tcp*[f]{binary assessment of compliance}\\
            return $k$, $l$
        }        
    \end{small}
\end{algorithm}

\section{Experiments}
This section describes how we experiment with the system based upon parameters identified in Eq. \ref{eq:avg_assessment}.

\vspace{1mm}
\noindent \textbf{Query $(X, Y)$ and benchmark programs $V$} -- The objective of the assessment process is to check whether a pair of query SWCs $(X, Y)$ complies with the average realization of the CH pattern seen in a separate set $V$ of known instances. With the help of architects who are familiar with the TAS corpus, we first identify $\mathbf{21}$ known instances of the CH pattern and curate them into a set $\mathcal{V}$. Next, we design a controlled experiment by selecting two types of queries. \looseness=-1
\begin{itemize}[wide, labelindent=0pt]
    \item \emph{The positive query} -- where the query $Q^+ \in \mathcal{V}$ is known to be an implementation of the CH pattern that is likely to satisfy the condition specified in Eq.\ref{eq:avg_offset}. The benchmark set in this case is $V = \mathcal{V} \setminus \{Q^+\}$, which is all known instances of the pattern excluding the instance chosen as the test input.
    \item \emph{The negative query} -- where the query $Q^- \in \mathcal{Q} \setminus \mathcal{V}$ is known to not implement the CH pattern and is therefore unlikely to satisfy Eq.\ref{eq:avg_offset}. Here, the benchmark set $V = \mathcal{V}$ includes all known instances of the CH pattern in the TAS corpus. Since we expect negative queries to perform poorly during the assessment, they help establish a baseline for the evaluating the accuracy of the assessment process. \looseness=-1
\end{itemize}

Consider a pair of SWCs $(C, H) \in \mathcal{V}$, that is known to implement the CH pattern. While it is most straightforward to implement each SWC in the pair as one program, this is not always practical. As shown in Figure \ref{fig:swc-app}a, some SWCs include a lot of functionality in which case it is necessary to split its code into several programs or files. Practically, therefore, the SWCs are of the form $C = \{C_1, C_2, ..., C_M\}$ and $H = \{H_1, H_2, ..., H_N\}$, each of them being implemented using multiple programs. This complicates the assessment process since the system $\mathcal{S}$ is designed only to handle a pair of programs and not a pair of sets. A simple way to circumvent this limitation is to `unroll' the set $\mathcal{V}$ into a Cartesian product set as follows. \looseness=-1
\begin{equation}
    \label{eq:cartesian_product}
    \begin{aligned}
        \mathcal{V}^* = \{(c, h): ~c \in C, h \in H \ : ~(C, H) \in \mathcal{V}\}
    \end{aligned}    
\end{equation}

For every known instance of the CH pattern $(C, H) \in \mathcal{V}$, the product set $\mathcal{V}^*$ pairs each program in the controller SWC $C$ with every program in the handler component $H$. This process results in a total of $\mathbf{63}$ pairs, which we use as likely queries in our experiments. By drawing queries $Q^+ \in \mathcal{V}^*$, the advantage is that we exhaustively present all combinations in a paired form that is suitable for assessment using Eq.\ref{eq:avg_assessment}. The disadvantage is that even if at the component level every pair $(C, H) \in \mathcal{V}$ is a known instance of the CH pattern, not every pair $(c, h) \in \mathcal{V}^*$ at the program level is a `true' controller-handler pair that implements elementary aspects of the CH pattern. Considering that several instances of the CH pattern are implemented using multiple programs, and that the assessment system is currently designed to work only with a pair of programs, we accept the risk and loosen the definition of the CH pattern. Every pair in the product set $\mathcal{V}^*$ is considered as a true pair and is presented as a positive case for testing, while also being used to calculate $r$. Negative queries $Q^-$ are simply drawn by picking two random programs from TAS as long as neither of them appear in $\mathcal{V}^*$. \looseness=-1

\vspace{1mm}
\noindent \textbf{The PLM $\mathcal{F}$} -- As the heart of the automated compliance assessment system, the neural PLM $\mathcal{F}$ can be seen as the machine counterpart of a human architect who conducts the same assessment manually. With such an analogy, we now reason about the level of information with which $\mathcal{F}$ is trained and its relation to the quality of assessment. The model pre-trained using Eq.\ref{eq:obj-mlm} on a C-language corpus extracted from GitHub -- which we now denote as $\mathcal{F}_A$ -- is a C-programming expert. Using this model is akin to asking a human expert in C-programming, but one who has no experience in automotive application design and development, to assess compliance with the CH pattern. While it is not impossible for such an expert to conduct this assessment, it is reasonable that an awareness of relevant domain and design concepts would ease the process. To a C-programming expert, we contend that such awareness can be introduced in three stages. The first stage would be to increase awareness about the automotive-domain, i.e. the pattern of token usage (its naturalness) in its application code. Second comes design-related knowledge, mainly the concept of SWCs, which is fundamental to the definition of the CH pattern. Third, would be the concept of controllers and handlers, the subjects of assessment. Like \cite{DBLP:conf/acl/GururanganMSLBD20}, we achieve the first stage -- improving domain-familiarity -- by simply continuing to pre-train $\mathcal{F}_A$ on code from TAS. The second stage requires inducing the knowledge of a SWC -- a set of programs that jointly realize functionality. We do this by first assembling a set $C = \{(A, P, N)\}_{i=1}^M$ of programs from TAS, such that $A$ and $P$ belong to the same SWC, while $N$ belongs to a different SWC. Then, we use the triplet loss to cluster embeddings of programs that belong to a SWC, while keeping those of programs from different SWCs further apart. To simultaneously ensure that this SWC-based clustering does not majorly disrupt the embedding geometry, and to impart domain familiarity, we combine the MR task on the TAS corpus with SWC-clustering as shown below. \looseness=-1
\begin{equation}
    \label{eq:swc-clustering}
    \begin{aligned}
        \mathcal{F}_B = \argmin_{\mathcal{F}} \mathbb{E}_{(A, P, N) \in C} &~~TR(A, P, N; ~\mathcal{F}) + MR(A; ~\mathcal{F}) \\
        TR(A, P, N; ~\mathcal{F}) &= (|| e_A - e_P||^2 - || e_A - e_N||^2)
    \end{aligned}    
\end{equation}

The resulting fine-tuned model $\mathcal{F}_B$ is thus more familiar with domain and design concepts related TAS in comparison to $\mathcal{F}_A$. For the third stage of inducing knowledge about controller and handler programs, we follow a similar approach of encouraging the PLM to respectively cluster these programs by type. To achieve this, we assemble (1) a set $D_C = \{(C_1, C_2, A)\}_{i=1}^M$ with $C_1$ and $C_2$ being controllers and $A$ being a non-controller program from the TAS corpus, and (2) a set $D_H = \{(H_1, H_2, B)\}_{i=1}^N$, with $H_1$ and $H_2$ being handler programs and $B$ being a non-handler program. We then fine-tune $\mathcal{F}_B$ using the triplet loss on the combined set $D = D_C \cup D_H$, resulting in a model $\mathcal{F}_C$ that is aware of the concept of controllers and handlers. \looseness=-1
\begin{equation}
    \label{eq:ch-clustering}
    \begin{aligned}
        \mathcal{F}_C = \argmin_{\mathcal{F}} \mathbb{E}_{(A, P, N) \in D} &~~TR(A, P, N; ~\mathcal{F}) + MR(A; ~\mathcal{F}) 
    \end{aligned}    
\end{equation}

By assessing design compliance using models $\mathcal{F}_A$, $\mathcal{F}_B$, and $\mathcal{F}_C$, respectively representing increasing awareness of concepts relevant to the assessment, we analyze the influence of such awareness. This assessment is conducted on an equal number of positive ($Q^+$) and negative ($Q^-$) queries. For each query, results are collected in terms of a discrete rank and a binary label (see Procedure \ref{alg:compliance}).  \looseness=-1

\section{Results}
The primary tool which we use for analyzing the results are the labels $l$ collected for each query. This binary label indicates whether the query has been evaluated by the system $\mathcal{S}$ to comply with or deviate from the CH pattern. The controlled experiment using positive and negative queries, which are known to comply and deviate from the pattern, allows collection of results of each of these cases into lists $L^+$ and $L^-$ respectively. Thus, true positive (TP) assessments are those labels in $L^+$ that evaluate to \texttt{True} and false negatives (FN) are those that evaluate to \texttt{False}. False positive (FP) and true negative (TN) assessments are similarly identifiable from $L^-$, as shown below. \looseness=-1

\begin{equation}
    \label{eq:conf-matrix}
    \begin{aligned}
        TP: \{l ~| ~l==True, l\in L^+\} ~~ FN: \{l ~| ~l==False, l\in L^+\}\\
        FP: \{l ~| ~l==True, l\in L^-\} ~~ TN: \{l ~| ~l==False, l\in L^-\}\\
    \end{aligned}    
\end{equation}

Using this, we build the confusion matrix (Table \ref{table:conf-matrix}) and performance metrics of the assessment process (Table \ref{table:pred-perf}). These metrics help us answer the research questions posed in our problem statement.\looseness=-1

\begin{table}[h]
    \centering
    \caption{Compliance assessment -- confusion matrix$^{1, 2}$}
    \resizebox{85mm}{!}{
    \begin{threeparttable}
        \begin{tabular}{| c| c| c| c| c| c| c|}
            \hline  
            \multirow{2}{*}{\backslashbox{Queries}{Prediction ($l$)}}
            & \multicolumn{2}{c|}{$\mathcal{F}_A$}
            & \multicolumn{2}{c|}{$\mathcal{F}_B$}
            & \multicolumn{2}{c|}{$\mathcal{F}_C$} \\ 
            \cline{2-7}
            \rule{0pt}{3ex}            
            & \textbf{True} & \textbf{False} & \textbf{True} & \textbf{False} & \textbf{True} & \textbf{False} \\
            \hline \textbf{Positive ($Q^+$) - 63} & 
            22 (0.35) &
            41 (0.65) &
            37 (0.59) &
            26 (0.41) &
            50 (0.80) &
            13 (0.20) \\ [1ex] 
            \hline \textbf{Negative ($Q^-$) - 63} & 
            ~8 (0.13) &
            55 (0.87) &
            ~7 (0.11) &
            56 (0.89) &
            4  (0.06) &
            59 (0.94) \\ [1ex] \hline
        \end{tabular}
        \begin{tablenotes}[flushleft]\normalsize
            \item[1] Confusion matrix on labels $L^+$ and $L^-$ calculated according to Eq.\ref{eq:conf-matrix}
            \item[2] For definition of each label $l\in L^+ or L^-$ refer to Procedure \ref{alg:compliance}
        \end{tablenotes}
    \end{threeparttable}
    }
    \label{table:conf-matrix}    
\end{table}

\begin{table}[h]
    \centering
    \caption{Compliance assessment -- performance metrics}
    \begin{threeparttable}
        \begin{tabular}{| c| c| c| c|}
            \hline \textbf{Metric} & \textbf{with $\mathcal{F}_A$} & \textbf{with $\mathcal{F}_B$} & \textbf{with $\mathcal{F}_C$} \\ [0.5ex] 
            \hline Accuracy & 0.611 & 0.738 & 0.860\\ [0.5ex] 
            \hline Recall & 0.349 & 0.587 & 0.790\\ [0.5ex]
            \hline Precision & 0.733 & 0.840 & 0.920\\ [0.5ex]            
            \hline F1 score & 0.473 & 0.691 & 0.850\\ [0.5ex] 
            \hline
        \end{tabular}
    \end{threeparttable}
    \label{table:pred-perf}    
\end{table}

\vspace{1mm}
\noindent \textbf{RQ1: assessing design compliance using neural PLMs} -- Performance metrics in Table \ref{table:pred-perf} show encouraging signs that a system for assessing compliance of programs $(X, Y)$ with the CH design pattern can be constructed using a neural language model trained on nothing but source code. Even with the model $\mathcal{F}_A$, which is pre-trained purely on non-automotive code, the system is capable of identifying instances of the CH pattern with a precision of more than $0.70$. As also seen in Table \ref{table:conf-matrix}, with a high True Negative Rate (TNR) ($0.87$), the system is particularly adept at correctly identifying non-compliant instances of the pattern. The main concern, seen from the same table, is of course the very high False Negative Rate (FNR) of $0.65$. That is, the system built using $\mathcal{F}_A$ is misclassifying a majority of known instances of the CH pattern as non-compliant. The high FNR, in turn, lowers the accuracy, precision and F1 score. Thus, while the performance of design compliance assessment using $\mathcal{F}_A$ is encouraging, it remains unsatisfactory. We reason that there are three main factors that could explain the high FNR. The first is the product set $\mathcal{V}^*$, which considers all possible pairs of programs from applications that are known instances of the CH pattern. The introduction of doubtful pairs could taint both the average jointness benchmark $r$ and whether a positive test input is genuinely so. The second reason could be the lack of familiarity with TAS domain and design in $\mathcal{F}_A$, due to which program embeddings are arranged in such a way that the benchmark vector $r$ does not serve as a good offset. The third reason could be some weakness in assessment using the average jointness benchmark $r$. Results from testing with models $\mathcal{F}_B$ and $\mathcal{F}_C$ shows that it is less likely to be due to a weakness in the assessment approach. \looseness=-1

\vspace{1mm}
\noindent \textbf{RQ2: assessment using PLMs with increased knowledge} -- Having been pre-trained only using the GitHub corpus, one weakness in $\mathcal{F}_A$ is that it is less aware of domain and design-related specializations in the TAS corpus. This is precisely why we train models $\mathcal{F}_B$ and $\mathcal{F}_C$ by explicitly providing this information. Assessment using $\mathcal{F}_B$, which learns domain-specific naturalness and the concept of SWCs used in the TAS corpus, leads to a strong reduction of the FNR to $0.41$. The consequent improvement in the F1 score to $0.7$ is also noteworthy. This clearly indicates that inducing the knowledge of SWCs directly leads to an improvement in the quality of assessment. Using model $\mathcal{F}_C$ -- which is trained to understand controller and handler programs -- for the assessment leads to yet another strong reduction in the FNR to $0.2$, due to which the precision and F1 score commendably increase to $0.92$ and $0.85$ respectively. The clustering objectives (Eqs. \ref{eq:swc-clustering} and \ref{eq:ch-clustering}), are therefore likely to have resulted in an arrangement of embeddings that better satisfies Eq. \ref{eq:avg_offset}. These observations clearly indicate that using a PLM with an increased level of awareness about the domain and its design, results in a much more accurate assessment. Even with a marked improvement in the quality of assessment, the FNR remains a concern. To analyze this, there is a need to go beyond binary assessment to a finer method. \looseness=-1

\begin{figure}[h]
    \centering
    \includegraphics[width=0.9\linewidth]{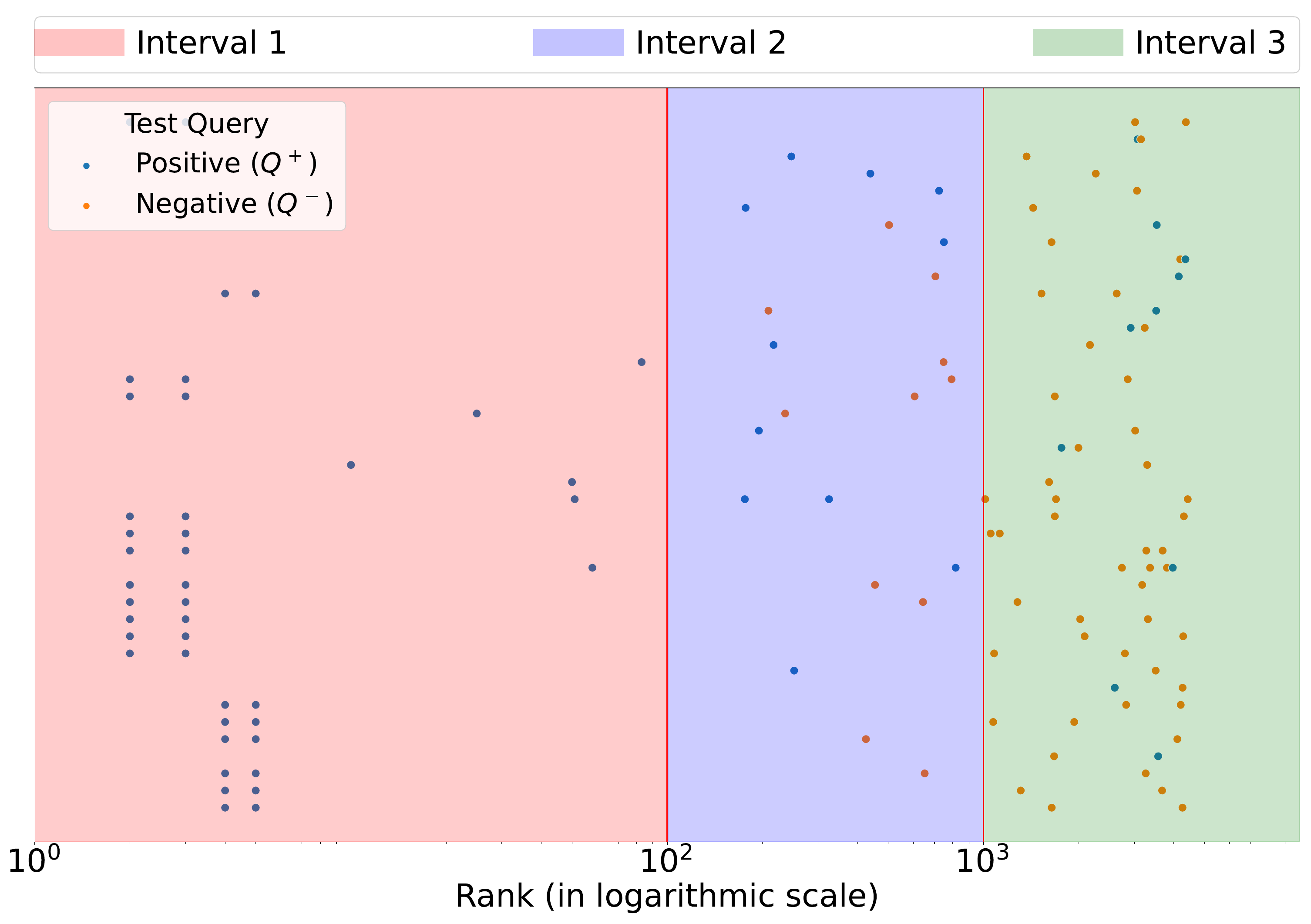}
    \caption{Calibrating the results of assessment into interpretable intervals using expert review}
    \label{fig:rank-dist}
\end{figure}

\vspace{1mm}
\noindent \textbf{RQ3: easing interpretation of assessment} -- Analyzing the binary labels of compliance ($L^+$ and $L^-$), using the confusion matrix and metrics derived from it, helps evaluate the performance of the assessment system. While this is necessary to build confidence in the system, from the perspective of an architect or developer, it is equally important to understand \emph{why} the system assesses a query as complying or deviating from the CH pattern. Since this requires much more nuance than a binary label, we turn to the rank $k$ to gain a finer interpretation of the assessment. Specifically, we analyze the distribution of $K^+$ and $K^-$ of ranks respectively collected for positive and negative queries. For brevity, we confine our analysis to the best performing system that uses the model $\mathcal{F}_C$. First we begin by visualizing the spread of ranks shown in Figure \ref{fig:rank-dist}. Inspecting the spread of ranks for the positive cases $K^+$, allows us to demarcate three intervals of ranks where results cluster. Next, we sample queries from each interval and have them assessed by architects who are familiar with TAS. The manual assessment of sampled queries follows an approach similar to the one described in Section \ref{sec:compliance}. Using expert review we calibrate the results of the PLM-based compliance assessment system within each interval as described below. \looseness=-1

\begin{itemize}[wide, labelindent=0pt]
    \item \emph{Interval 1 (ranks 1-100)} -- The interval where a majority of positive cases cluster, it consists mostly of queries that are assessed by architects to be good implementations of the CH design pattern. Some are even judged to be textbook cases with the right interface and responsibility split between controller and handler programs. The best ranking instances in this interval are also those which exhibit bidirectional exchange of information between the two programs. The exchange follows standard practice of using the AUTOSAR Runtime Environment (RTE), seen in their use of \texttt{RTE\_read} and \texttt{RTE\_write} methods\footnote{Refer to roofHatch in released code for an example}. Cases that perform relatively worse within this interval (rank close to 100) are observed to implement unidirectional interaction, where the controller only reads from the handler, which is perfectly legitimate. Therefore, expert review generally considers test inputs that rank in this interval to be compliant with the CH pattern, with no need for refactoring. This is further strengthened by the fact that not a single negative test case is ranked by the system as being in this interval.\looseness=-1
    \item \emph{Interval 2 (ranks 100-1000)} -- Expert review indicates that positive queries in this interval show subtle deviations from the standard implementation of the CH pattern. One deviation is that, while the responsibility split is correct, the controller and handler programs do not interact directly with each other. The actual interaction, in this case, usually happens through some other program in the controller SWC, which is excluded due to the constraint that the system operates only on pairs of programs. This is, therefore, not a genuine violation and results simply due to a limitation in the system. More significantly, the other observed deviation is where there is direct interaction, but it does not take place through the AUTOSAR RTE. This is a subtle deviation which could benefit from refactoring. The fact that the assessment system consistently places such cases in the second interval is an encouraging observation. However, the deviations observed by expert review in this interval also seem to be characteristics observable in pairs of programs that are not controllers or handlers. For instance, it is plausible that random sampling from the relatively small TAS corpus results in a pair of non-interacting programs, one of which contains some application-like code and the other containing some code related to hardware. This could explain why some negative cases end up being ranked in this interval. In general, when the system ranks a query in this interval, it could be a candidate for refactoring. However, it is best if the automated assessment is manually verified to ensure that it is a genuine case.
    
    \item \emph{Interval 3 (ranks 1000-5000)} -- Very few positive cases rank in this interval. In some cases, the test input implements diagnostic routines, and not application logic. In others, the controller program is very small, containing only a few lines of code. Generally, therefore positive cases seem to rank in this interval because they are marked outliers compared to the average CH implementation. A cause for concern are the handful of cases which are genuine false negatives and are, in fact, assessed to be good implementations of the CH pattern. Moreover, a query ranked in this interval seems to deviate from the average implementation to such an extent that it is barely distinguishable from random queries drawn from TAS. A result in this interval therefore requires manual review by an expert.
\end{itemize} 

Thus, the greatest advantage of the system is its ability to identify true compliance with the CH pattern. Such cases, as verified by experts, rank in the first interval. Also, its tendency to rank subtle variations -- possible candidates for refactoring -- in the second interval shows its ability make nuanced judgments. Finding such deviations is a strong indicator of its practical utility. The inconclusive nature of results in the third interval, and the presence of some negative cases in the second, indicate the boundaries of this process. \looseness=-1

\vspace{1mm}
\noindent \textbf{Overall observations} -- First, queries that fall within the first two intervals are remarkably similar in character, meaning that observations apply quite consistently to cases within a given interval. This reflects the consistency of automated assessment using the average jointness benchmark. Second, this consistency eases practical use because when a query ranks within an interval, we have a reasonably good idea why this happens. This means that any subsequent design intervention can be precisely targeted to rectify suspected deviations. Third, the calibration process makes it possible to decide the conditions under which it is necessary for an architect to intervene. Ranks in the first interval do not require human verification, while those in the second and (especially) third intervals need active intervention. These observations thus point to the ingredients of a protocol for interpreting the results and, thus, practically using the system. However, we also observe a few caveats in the process which we now list. First, under the current process, the benchmark $r$ needs to be recalculated whenever there is a new instance of the CH pattern. Since this is not a computationally heavy process, we do not rate this as a major concern. An alternative would be to fix `golden' instances of the pattern so that the benchmark $r$ is itself fixed. While choosing such instances, it is important to ensure that legitimate variations are included. It would also be necessary to periodically audit the golden instances to ensure that they are up-to-date with the latest understanding of the pattern. Second, the ranking process depends upon all programs in the TAS corpus, meaning that the addition of new programs needs a recalibration of the results. In the worst case, the inclusion of a new set of highly specialized programs could severely disrupt the calibration. However, it is important to note that such risks are inherent to any benchmark that is derived from an evolving corpus. Third, there is a need to better understand the relationship between pattern compliance and rank. Consider the test input with a rank close to 100 (and thus in interval 1), but deviates from the textbook implementation because here the controller only reads from, and does not write to, the handler. Such a deviation seems sufficient for $\sim$100 programs in the TAS corpus to come in between the predicted and actual handler embeddings. While the empirical calibration process allows us to circumvent this, it is essential to understand the nature of intervening program embeddings. This is an investigation that we prioritize for future work. Overall, results from this study demonstrate a promising method to construct an automated system for measuring design pattern compliance using neural language models trained on source code. \looseness=-1

\section{Discussion}
\label{sec:discussion}
\vspace{1mm}
\noindent \textbf{Relationship with embedding regularities} -- The central idea of our PLM-based system for design compliance, captured in Eq.\ref{eq:avg_offset}, is whether the representation of average jointness $r$ serves as an effective offset vector for query embeddings $(e_X, e_Y)$. For this condition to hold across several possible queries, the embeddings of controller and handler programs need to follow a specific pattern of arrangement. The geometrical relationship between the embeddings of semantically related entities has been the focus of extensive study in neural natural language processing. Most famously, \cite{DBLP:conf/naacl/MikolovYZ13} studied regularities in the embeddings of word pairs that are associated by a similar concept. Using pairs of words $(Man, Woman)$ and $(King, Queen)$, the word2vec language model has been shown to learn embeddings such that $e_{King}-e_{Man}+e_{Woman} \approx e_{Queen}$. If the model has a proper understanding of the analogical relationship between these pairs of words then, as shown by \cite{DBLP:conf/naacl/MikolovYZ13}, these four word embeddings approximate a parallelogram. Building upon this idea, \cite{DBLP:conf/coling/ZhuM20} showed that embeddings of pairs of sentences that are related by the same concept show a similar parallelogram geometry. Our entire approach can be reinterpreted as examining the embedding regularities of pairs $(X, Y)$ of programs that are related by the same concept -- the CH design pattern. If the neural PLM $\mathcal{F}$ used in the assessment system correctly encodes the jointness that underlies the CH design pattern, embeddings of pairs of programs $(C_1, H_1)$ and $(C_2, H_2)$ that implement this pattern should approximate a parallelogram. In which case, $e_{C_2} + (e_{H_1} - e_{C_1}) \approx e_{H_2}$ must hold, which is a special case of the average jointness benchmark with one known pattern instance. If this parallelogram geometry consistently holds across several instances of the pattern, the average jointness vector $r$ naturally serves as an effective offset between the program embeddings of any given instance $(X, Y)$. Further, since regularity is essential for compliance using $r$ as the offset, we reason that clustering objectives (Eqs.\ref{eq:swc-clustering} and \ref{eq:ch-clustering}) strengthens it, improving the quality of the assessment process. Additionally, \cite{DBLP:conf/coling/DrozdGM16} formalized the idea of testing the regularity of one pair of related words using the average offset of other pairs of similarly related words -- a technique that they refer to as 3CosAvg. Their use of the average offset closely reflects our construction of the average jointness $r$ as the benchmark for assessment. The fact that the system we design for design compliance assessment is firmly grounded in extensively studied properties of neural language models, inspires further confidence in our approach. \looseness=-1

\vspace{1mm}
\noindent \textbf{The quality of program embeddings} -- The system we construct for assessing compliance with a design pattern is built upon program embeddings, which are vector representations of programs extracted from the PLM $\mathcal{F}$. The quality of the assessment process is therefore highly dependent upon the quality of the representation. Among the factors that influence this quality, perhaps the most important is the objective that is used to train the model. PLMs used in our study are primarily trained using the masked reconstruction task shown in Eq.\ref{eq:obj-mlm}. The simplicity of the MR task is undoubtedly its key advantage. However, a major shortcoming of the BERT masking recipe is that, by uniformly choosing 15\% of the tokens to be masked, only tokens that are numerically abundant -- but semantically less significant (like \texttt{;}) -- are more likely to be masked. In order to successfully reconstruct a token like \texttt{;} it is often sufficient to simply learn concepts in a local scope, like the likelihood of the end of a statement. Thus, with the model rarely being tasked with reconstructing tokens that are semantically significant, it is less equipped to learn global concepts like design. This could explain why the base model $\mathcal{F}_A$, which is pre-trained only using MR, performs worst. This weakness of MR is well-documented in literature and several interesting alternatives have been proposed that encourage the model to learn more global concepts. One option is to modify the masking recipe like \cite{DBLP:journals/corr/abs-1904-09223}, which masks selected phrases and \cite{DBLP:journals/tacl/JoshiCLWZL20}, which masks larger spans of tokens. Another option is to use \cite{DBLP:conf/iclr/ClarkLLM20} and \cite{DBLP:conf/acl/LewisLGGMLSZ20}, which task a model to detect replaced, permuted, inserted or deleted tokens. As tasks that are more complex than reconstructing simple tokens, they encourage the model to gain a deeper understanding of program contexts. Another interesting alternative class of training objectives are those that selectively obfuscate tokens. For instance, a de-obfuscation objective proposed by \cite{DBLP:journals/corr/abs-2102-07492} obfuscates class, method, and variable names before tasking the model to recover them. Since the successful completion of this task requires a deeper and broader understanding of the program, they may lead to embeddings that are better suited for a design assessment. While we reason the fine-tuning objectives that improve domain and design-related awareness (Eqs.\ref{eq:swc-clustering} and \ref{eq:ch-clustering}) are likely to remain important, setting a task that is more complex than MR may result in a much more powerful base model $\mathcal{F}_A$. We leave this investigation for future work. \looseness=-1

\vspace{1mm}
\noindent \textbf{Training beyond code} -- Our results clearly show that it is possible to construct a system for assessing design compliance using PLMs trained on source code. However, we do not necessarily advocate a code-only training approach for imparting design knowledge. In addition to source code, automotive software engineering, which follows the AUTOSAR standard, captures additional engineering information using the standard ARXML modeling language. From the perspective of design awareness, would it therefore be helpful to explicitly train PLMs with ARXML models? The answer depends, of course, upon whether such models provide additional design awareness. If most of the information in ARXML models is likely to be replicated in code, then using them for training is unlikely to enhance design understanding. Otherwise, if design models do contain some information not discernible in code, it may indeed be helpful to additionally train with such information. Assessing the usefulness of engineering information in ARXML for design compliance assessment is an investigation that we leave for future work. \looseness=-1
\section{Related work}
To the best of our knowledge, our work is the first attempt to apply neural language models for measuring design compliance. In software engineering, our work closely relates with the task of design pattern detection. A recent survey of this area \cite{DBLP:journals/air/YarahmadiH20} reveals that around 20\% of reported methods take a machine learning approach, mostly using classical algorithms. Examples include \cite{thaller2019feature} which compares pattern instances by modeling them as graphs, and \cite{oberhauser2020machine} and \cite{NAZAR2022111179} which use artificial neural network and random forest models respectively to classify pattern instances. We reason that the key advantage of our use of neural language models is the level of nuance that it can apply for judging design. A BERT-like PLM, which has been shown to learn nuanced contextual information, could be vital for assessing design, where firm judgments are rare. Also, unlike the majority focus on pattern detection, we develop a technique for measuring compliance with a given pattern, including steps to identify the source of deviation. Moreover, our study focusing upon embedded control systems would also be a useful addition to an area that mostly focuses on object-oriented design.\looseness=-1

\vspace{1mm}
As discussed in detail in Section \ref{sec:discussion}, our approach for compliance assessment closely relates to the property of linguistic regularity observed in neural natural language models \cite{DBLP:conf/naacl/MikolovYZ13}. Most experiments, as surveyed in \cite{bakarov2018survey}, study this property as a way to evaluate the quality of word embedding models. Few of them apply this property in a predictive setting by framing an analogy completion task where, given a triplet $(A, B, C)$, they predict $D$ such that $(A, B)$ and $(C, D)$ are analogical pairs. Studies \cite{grave2018learning} and \cite{lim2021classifying} approach this task respectively using popular word2vec and GloVe embedding models, while \cite{da2020generating} uses sense embeddings derived from word2vec. An example of the property being studied in a specialist domain is \cite{chen2021domain} which fine-tunes GloVe on a corpus related to radiology, and uses its embeddings for the analogy completion task. Similar to our departure from word embedding models, \cite{ushio2021bert} studies this property in pre-trained contextual neural language models. The work we survey can therefore be seen to relate to parts of our assessment system, but we build a pipeline that not only analyzes embedding regularity, but also interprets it within the context of software and its design. In doing so, we also tie the property of embedding regularities to a concrete application.\looseness=-1

\section{Conclusions}
This work demonstrates how neural language models trained on source code can be used to measure whether a set of programs comply with desired design properties. Compliance is measured by inspecting the geometrical properties -- specifically the regularity -- of query program embeddings. Our work also includes techniques to significantly improve the accuracy of the assessment by explicitly providing the model with domain and design-related information. Experiments performed on an automotive code corpus result in a prediction precision of $92\%$. We also present how the model predictions can be incorporated into a design review methodology in order to provide valuable feedback to automotive software architects.
\looseness=-1

\begin{acks}
This work is partially funded by Chalmers AI Research Center (CHAIR) project T4AI.
\end{acks}

\bibliographystyle{ACM-Reference-Format}
\bibliography{bibliography}



\end{document}